\documentclass[pdflatex,sn-mathphys-num]{sn-jnl}

\usepackage{graphicx}%
\usepackage{multirow}%
\usepackage{amsmath,amssymb,amsfonts}%
\usepackage{amsthm}%
\usepackage{mathrsfs}%
\usepackage[title]{appendix}%
\usepackage{xcolor}%
\usepackage{textcomp}%
\usepackage{manyfoot}%
\usepackage{booktabs}%
\usepackage{algorithm}%
\usepackage{algorithmicx}%
\usepackage{algpseudocode}%
\usepackage{listings}%

\theoremstyle{thmstyleone}%
\theoremstyle{thmstyletwo}%

\theoremstyle{thmstylethree}%

\begin{document}

\title[Article Title]{Super-Moiré Spin Textures in Twisted Antiferromagnets}


\author[1]{\fnm{King Cho} \sur{Wong}}
\equalcont{These authors contributed equally to this work.}

\author*[1]{\fnm{Ruoming} \sur{Peng}}\email{ruoming.peng@pi3.uni-stuttgart.de}
\equalcont{These authors contributed equally to this work.}

\author[2]{\fnm{Eric} \sur{Anderson}}
\equalcont{These authors contributed equally to this work.}

\author[3]{\fnm{Jackson} \sur{Ross}}
\equalcont{These authors contributed equally to this work.}

\author[4]{\fnm{Bowen} \sur{Yang}}

\author[4]{\fnm{Meixin} \sur{Cheng}}

\author[1]{\fnm{Sreehari} \sur{Jayaram}}

\author[1]{\fnm{Malik} \sur{Lenger}}

\author[1]{\fnm{Xuankai} \sur{Zhou}}

\author[1]{\fnm{Yan Tung} \sur{Kong}}

\author[5]{\fnm{Takashi} \sur{Taniguchi}}

\author[6]{\fnm{Kenji} \sur{Watanabe}}

\author[7]{\fnm{Michael A.} \sur{McGuire}}

\author[1]{\fnm{Rainer} \sur{Stöhr}}

\author[4]{\fnm{Adam Wei} \sur{Tsen}}

\author*[3,8,9]{\fnm{Elton J.G.} \sur{Santos}}\email{esantos@exseed.ed.ac.uk}

\author[2,10]{\fnm{Xiaodong} \sur{Xu}}

\author[1,11]{\fnm{Jörg} \sur{Wrachtrup}}

\affil[1]{\orgdiv{3rd Physikalisches Institut}, \orgname{University of Stuttgart}, \city{Stuttgart}, \postcode{70563}, \country{Germany}}

\affil[2]{\orgdiv{Department of Physics}, \orgname{University of Washington}, \city{Seattle}, \postcode{98195}, \state{WA}, \country{USA}}

\affil[3]{\orgdiv{Institute for Condensed Matter Physics and Complex Systems}, \orgname{The University of Edinburgh}, \city{Edinburgh}, \postcode{EH9 3FD}, \country{United Kingdom}}

\affil[4]{\orgdiv{Department of Chemistry}, \orgname{University of Waterloo}, \city{Waterloo}, \postcode{ON N2L 3G1}, \country{Canada}}

\affil[5]{\orgdiv{Research Center for Materials Nanoarchitectonics}, \orgname{National Institute for Materials
Science}, \city{Tsukuba}, \postcode{305-0044}, \country{Japan}}

\affil[6]{\orgdiv{Research Center for Electronic and Optical Materials}, \orgname{National Institute for Materials
Science}, \city{Tsukuba}, \postcode{305-0044}, \country{Japan}}

\affil[7]{\orgdiv{Materials Science and Technology Division}, \orgname{Oak Ridge National Laboratory}, \postcode{TN 37830}, \country{USA}}

\affil[8]{\orgdiv{Higgs Centre for Theoretical Physics}, \orgname{ University of Edinburgh}, \city{Edinburgh}, \country{United Kingdom}}

\affil[9]{\orgdiv{Donostia International Physics Center}, \city{20018 Donostia-San Sebastián}, \country{Spain}}

\affil[10]{\orgdiv{Department of Materials Science and Engineering}, \orgname{University of Washington}, \city{Seattle}, \postcode{WA 98195}, \country{USA}}

\affil[11]{\orgdiv{Max Planck Institute for Solid State Research}, \city{Stuttgart}, \postcode{70569}, \country{Germany}}


\abstract{Stacking two-dimensional (2D) layered materials offers a powerful platform to engineer electronic and magnetic states. In general, the resulting states – such as Moiré magnetism - have a periodicity at the length scale of the Moiré unit cell. Here, we report a new type of magnetism - dubbed a super-Moiré magnetic state, which is characterized by long-range magnetic textures extending beyond the single Moiré unit cell - in twisted double bilayer chromium triiodide (tDB CrI$_3$). We found that at small twist angles, the size of the spontaneous magnetic texture increases with twist angle, opposite to the underlying Moiré periodicity. The spin-texture size reaches a maximum of about 300 nm in 1.1° twisted devices, an order of magnitude larger than the underlying Moiré wavelength, and vanishes at twist angles above 2°. Employing scanning quantum spin magnetometry, the obtained vector field maps suggest the formation of antiferromagnetic Neel-type skyrmions spanning multiple Moiré cells. The twist angle dependent study combined with large-scale atomistic simulations suggests that complex magnetic competition between the Dzyaloshinskii–Moriya interaction, magnetic anisotropy, and exchange interactions controlled by the relative rotation of the layers produces the topological textures which arise in the super-Moiré spin orders. }




\maketitle

\section*{Introduction}\label{sec1}

Controlling magnetic interactions in strongly correlated electronic systems has paved the way for designing exotic phases\cite{dietl2014dilute,wiesendanger2016nanoscale,hellman2017interface,mak2019probing}, including topological states such as skyrmions\cite{wiesendanger2016nanoscale,fert2017magnetic,jiang2017skyrmions}, merons\cite{gobel2021beyond}, and spiral spin orders\cite{tokura2010multiferroics}, as well as frustrated magnetic phases\cite{lacroix2011introduction} like spin glasses\cite{binder1986spin} and quantum spin liquids\cite{balents2010spin}. These developments have significant implications for understanding complex systems\cite{amit1985spin,moessner2001ising}, developing topological quantum computation\cite{freedman2003topological}, and are promising for novel spintronic applications\cite{hirohata2020review,chen2024all}. All these non-trivial states involve magnetic energy competition of multiple magnetic interactions, leading to unusual magnetic phenomena. It has been demonstrated in magnetic systems with coexisting ferromagnetic and antiferromagnetic interactions lead to strong competition\cite{furdyna1988diluted}, thereby frustrating the magnetic arrangement into spin glasses hosting skyrmions\cite{karube2018disordered,kurumaji2019skyrmion}. Moreover, interactions can be tailored by alternating the deposited materials to induce new magnetic states\cite{jiang2017skyrmions}, in contrast with conventional magnets. Recent advances in synthetic antiferromagnets\cite{duine2018synthetic}  have demonstrated that engineering the interlayer and interfacial interactions facilitates intriguing noncollinear and topological phases\cite{legrand2020room}.

Two-dimensional (2D) materials have opened new possibilities to control electronic and magnetic interactions\cite{gibertini2019magnetic,gong2019two,huang2020emergent}. In particular, stacking layered 2D materials with a small relative twist angle leads to the formation of Moiré superlattices, which can dramatically alter the electronic band structure and give rise to a wide range of emergent phenomena—including correlated insulating states, topological phases, superconductivity, and magnetism \cite{he2021moire,andrei2021marvels}. In twisted 2D magnetic bilayers, both theoretical and experimental studies have shown a correlation between the stacking order in a single Moiré cell and their magnetic responses\cite{sivadas2018stacking,song2021direct,xu2022coexisting,xiao2021magnetization}, leading to complex magnetic states\cite{tong2018skyrmions,akram2021skyrmions,zheng2023magnetic,kim2023emergence,xie2023evidence,cheng2023electrically,yang2024macroscopic}. Notably, in the previous report on near-zero degree twisted CrI$_3$ \cite{song2021direct}, scanning quantum microscopy was utilized to directly visualize Moiré magnetism, revealing stacking-dependent antiferromagnetic (AFM) and ferromagnetic (FM) domains in twisted bilayer and double trilayer CrI$_3$.

While previous theoretical and experimental studies have primarily focused on interactions confined within a single Moiré unit cell, this framework may no longer suffice when additional interactions are present. In multilayer Moiré heterostructures, for example, the interference of two distinct Moiré patterns can produce a secondary periodicity—known as a super-Moiré potential—with a characteristic length scale larger than either Moiré lattice \cite{wang2019composite,uri2023superconductivity,xie2024strong}. In twisted magnetic bilayers, magnetic textures are typically understood in terms of interlayer interactions confined within a single Moiré unit cell. However, additional contributions—such as intralayer exchange and interface-driven magnetic interactions—can strongly compete with the interlayer coupling\cite{gibertini2019magnetic,gong2019two,huang2020emergent}. These competing effects have been previously considered only within the Moiré unit cell, where the magnetic textures are assumed to strictly follow the underlying lattice modulation. However, as the twist angle increases, the Moiré unit cell becomes progressively smaller, forcing all magnetic interactions into a more confined region. In such a regime, it becomes conceivable that the magnetic texture will decouple from the small Moiré length scale as suggested in micromagnetics\cite{brown1968fundamental}, giving rise to magnetic supercells whose size deviates from that of the Moiré lattice.

Here, we employed nitrogen vacancy (NV) scanning microscopy to explore the nanoscale magnetic responses of devices at various twist angles ($\leq$2°). Notably, we found that the local magnetization patterns of our twisted devices do not reflect the Moiré stacking as suggested by earlier reports\cite{xie2023evidence,cheng2023electrically}. Instead, we identified super-Moiré antiferromagnetic orders with a large characteristic length scale of hundreds of nanometers, far exceeding the lattice Moiré length scale of a few tens of nanometers. These super-Moiré patterns are driven by complex magnetic competition, in contrast with previous reports, where the super-Moiré potentials arise from the interference between two distinct Moiré patterns in twisted multilayer heterostructures. More importantly, our observations suggest the presence of antiferromagnetic (AFM) skyrmions in tDB CrI$_3$ after a field-cooling process. The vector field maps provide strong evidence of Neel-type skyrmions, marking the first experimental observation of such topological textures in any twisted two-dimensional magnetic system to date.


\section*{Extended Ferromagnetic Domains}

\begin{figure}[h!]
  \centering
  \includegraphics[width=1\textwidth]{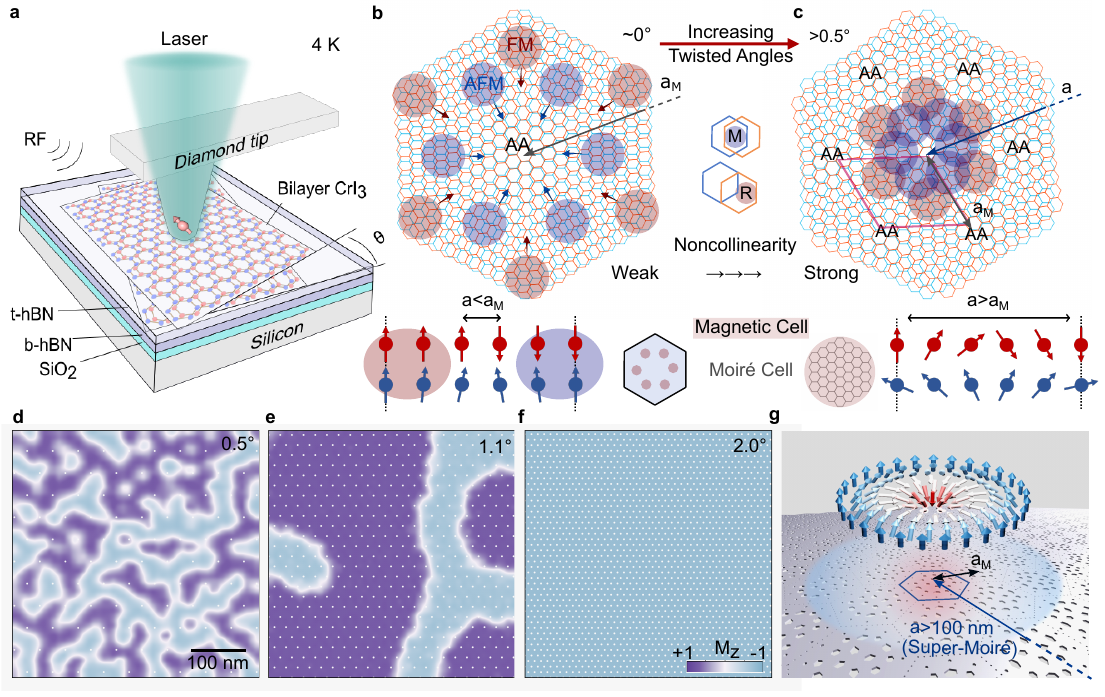}
  \caption{\textbf{a}. Schematic of the scanning quantum microscopy technique for the visualization of magnetic textures in tDB CrI$_3$. An NV center (red spin) is located at the apex of a diamond pillar. The NV is initialized by a green laser and controlled by a microwave signal. The sample device consists of two sheets of bilayer CrI$_3$ with a twist angle $\theta$ between them. \textbf{b}. Schematic of Moiré-modulated magnetic interactions at near-zero twist angle, showing FM regions (rhombohedral stacking, shaded red) and AFM regions (monoclinic stacking, shaded blue) are well separated. Thus, the magnetic texture closely follows the underlying Moiré lattice, forming sharp magnetic domain walls. In this regime, $a<a_M$, where $a$ and $a_M$ denote the length scale of magnetic texture and Moiré wavelength respectively. \textbf{c}. Schematic of competing magnetic orders at larger twist angles, where FM- and AFM-favored regions shift toward AA sites. Magnetic competition drives strong noncollinear spin textures, with magnetic domains extending beyond a single Moiré unit cell. In this regime, $a>a_M$. \textbf{d-f}. Atomistic simulated normalized magnetization maps over a 450-nm region for twist angles of 0.5°, 1.1°, and 2°. Only magnetic competition of $J_\perp$ and $J_\parallel$ is considered. A cross-section of the magnetization from a single layer of the tDB CrI$_3$ shows the merging of Moiré cells into larger magnetic textures with increasing twist angle. White dots mark the positions of underlying monoclinic sites for each twist angle. \textbf{g}. Schematic illustration of topological magnetic textures in relation to the underlying Moiré stacking. The arrows indicate the spin configuration in the 2D magnet, while the gray lattice beneath represents the Moiré superlattice. The shaded holes artistically depict regions of distinct stacking. A representative Moiré unit cell is outlined by a blue hexagon. The emergent super-Moiré magnetic textures span multiple Moiré unit cells, with characteristic length scales exceeding 100 nm.}
  \label{fig:figure1}
\end{figure}

Our magnetic systems consisted of two sheets of bilayer CrI$_3$ with small twist angles, as illustrated in Fig. \ref{fig:figure1}a. These twisted samples were fabricated using the "tear and stack" technique\cite{kim2016van} in an argon glovebox and were fully encapsulated with approximately 10 nm hexagonal boron nitride (hBN) flakes to prevent oxidation (More information about the sample devices on Supplementary Section 15-17). As a result of the twist, the local stacking of layers in tDB CrI$_3$ alternates between monoclinic and rhombohedral configurations, leading to periodically modulated magnetic couplings. Specifically, the inner two layers (layers 2 and 3) exhibit alternating ferromagnetic (FM, rhombohedral stacking) and antiferromagnetic (AFM, monoclinic stacking) coupling. Meanwhile, the interlayer interactions $J_\perp$ of the outer bilayers (e.g., layers 1 and 2; layers 3 and 4) are naturally antiferromagnetic. The intralayer interactions $J_\parallel$ of all four layers are expected to be ferromagnetic. At near-zero twist angles, the FM and AFM order regions are well separated with each other within the Moiré unit cell because the Moiré wavelength is large, as shown in Fig.~\ref{fig:figure1}b. In this regime, $a<a_M$, with $a$ and $a_M$ denoting the length scale of magnetic texture and Moiré wavelength, respectively. The magnetic texture at this small angle limit is well-described by a single-Moiré-cell model, with the magnetization strictly following the underlying Moiré superlattice stacking configuration and with sharp transitions between domains.

At a large twist angle, magnetic competition emerges as regions favoring interlayer FM and AFM coupling are brought closer together within a single Moiré cell, as shown in Fig. \ref{fig:figure1}c. In this regime, $a>a_M$. Therefore, a rich landscape of magnetic phases can arise featuring emergent magnetization patterns, extended domain walls, and noncollinear spin orders that span multiple Moiré cells. We performed large-scale atomic simulations of tDB CrI$_3$ with a record size of 450 nm, which consider only magnetic competition between $J_\perp$ and $J_\parallel$. At 0.5°, in one of the 4 layers, the simulations revealed large magnetic domains that span several Moiré unit cells {\color{black}(Fig.\ref{fig:figure1}d–f; see Supplementary Section 19 for simulation details).} As the twist angle is increased to 1.1°, these domains expanded, and at 2°, the magnetization becomes nearly uniform across the entire 450 nm simulation area. Such angle-dependent behavior is opposite to the predictions of the single Moiré model, which anticipates shrinking magnetic textures with increasing twist angles. Instead, strong magnetic competition stabilizes domains with sizes larger than the Moiré unit cell. Moreover, the presence of Dzyaloshinskii–Moriya interactions (DMI) and dipolar interactions induces canting in the spin configuration, resulting in extended magnetic textures. In particular, the uniform DMI induced by the hBN/CrI$_3$ interface promotes the formation of Néel-type structures with length scales above 100 nm. The interplay between these competing magnetic interactions is crucial for stabilizing noncollinear and topological magnetic structures that exceed the Moiré length scale, as illustrated in Fig.~\ref{fig:figure1}g.

To probe the underlying magnetic textures in tDB CrI$_3$, we utilized NV scanning microscopy (see Supplementary Section 1)\cite{maletinsky2012robust}. {\color{black}Previous reports have shown that this technique offers a sensitivity of a few $\mu\text{T}/\sqrt{\text{Hz}}$ and a spatial resolution limited only by the spacing between the NV center and the magnetic samples.}
Compared to conventional optical approaches, NV scanning microscopy offers significantly improved spatial resolution, and also allows direct reconstruction of local magnetization from detected stray field, as demonstrated in Fig. \ref{fig:figure2}a and \ref{fig:figure2}b. All measurements are performed at 4 K unless otherwise stated. From the magnetization maps of small-angle twisted ($\leq 2^\circ$) samples (magnetization reconstruction is discussed in Supplementary Section 2), we distinguish two categories of magnetic response in the twisted regions: FM domains with a magnetization of approximately $30~\mu_\text{B}/\text{nm}^2$ and AFM domains with a magnetization near $0~\mu_\text{B}/\text{nm}^2$ (see also Supplementary section 11). The uniform magnetization of $\sim30~\mu_\text{B}/\text{nm}^2$ across the majority of FM regions corresponds to a net two-layer magnetization of CrI$_3$\cite{song2021direct,huang2023revealing}. Both FM and AFM domains are strongly pinned by local defects and remain stable over multiple thermal cycles.

Despite the Moiré wavelengths at larger twist angles potentially being smaller than our scanning resolution of 30 nm, the detected FM domain magnetization significantly exceeds the expected maximum of $12~\mu_\text{B}/\text{nm}^2$ based on a Moiré unit cell 
model\cite{xie2023evidence,cheng2023electrically,yang2024macroscopic}. From our angle-dependent study (discussed in Supplementary Section 3 and 4), we find that the ratio of FM to AFM domains follows a similar trend to earlier RMCD measurements\cite{xie2023evidence}, exhibiting maximum FM response in samples twisted by 1.1°. We also notice a finite domain wall width between FM and AFM domains that is well above the spatial resolution of 30 nm, contrasting with the narrow domain wall width of below 10 nm expected in the single Moiré model. The domain wall width also exhibits clear twist angle dependence, as illustrated in Fig. \ref{fig:figure2}c. We identified the largest domain wall width of $\approx$118 nm in 1.1° twisted samples, which are predicted to exhibit the largest FM responses and noncollinear features. The domain wall length far exceeds the Moiré wavelength of the corresponding twist angle, which suggests that magnetic domains cannot be simply relaxed within a single Moiré cell. This is also captured in our atomic spin simulations, where no magnetic texture exists within a single Moiré cell at 1.1°. 

\section*{Emergence of {\color{black}Long-range} Antiferromagnetic Textures}

\begin{figure}[h!]
  \centering
  \includegraphics[width=1\textwidth]{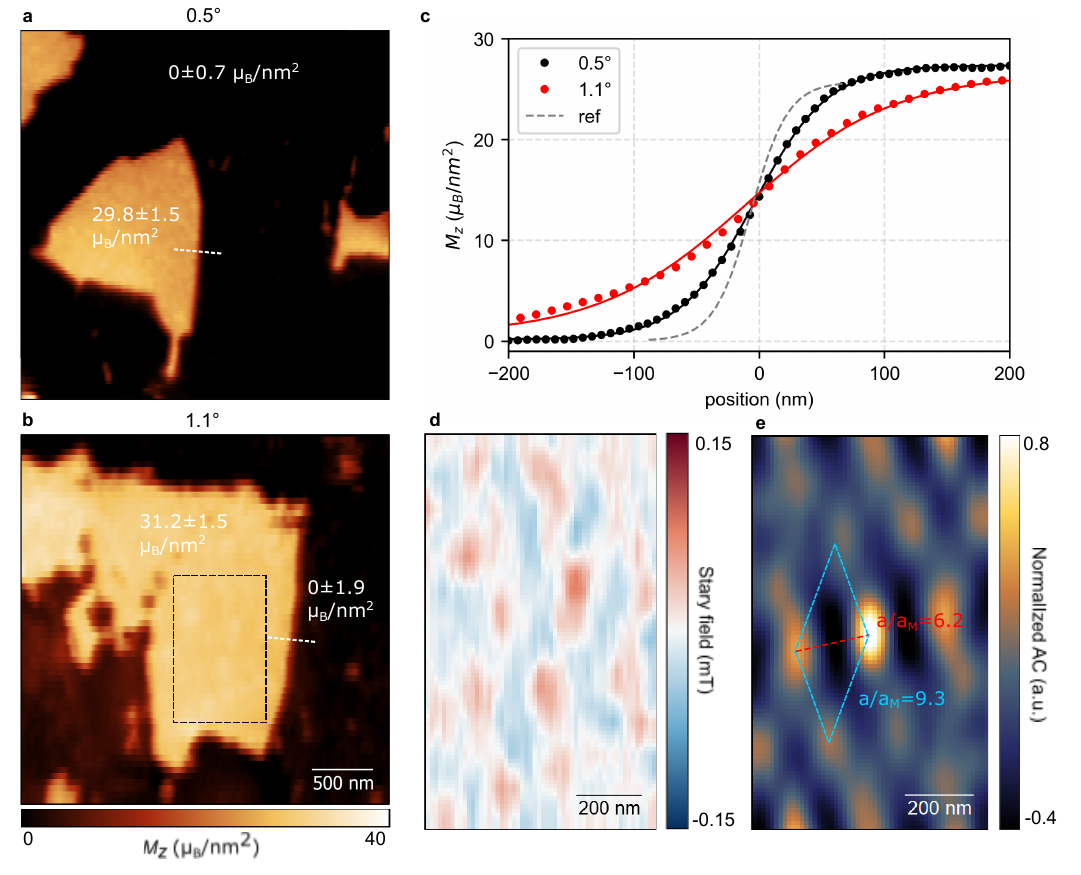}
  \caption{\textbf{a, b}. 2D magnetization maps of tDB CrI$_3$ at 0.5° and 1.1° twist angles, showing randomly distributed FM and AFM domains. The FM domains have a magnetization of around 30 $\mu_\text{B}/\text{nm}^2$, while other regions are all nearly 0 $\mu_\text{B}/\text{nm}^2$. \textbf{c}. Representative FM domain wall linecuts of 0.5° (red dots) and 1.1° (black dots) of tDB CrI$_3$ samples, fitted with a hyperbolic tangent function (red and black lines). The fitting for 0.5° and 1.1° linecuts showed domain wall widths of 58.5 nm and 118.2 nm. Positions of the linecuts are denoted on (a,b) as white dashed lines. \textcolor{black}{A domain wall linecut from a twisted bilayer CrI$_3$ (dashed grey line) serves as a reference for minimum resolvable domain wall width of around 30 nm. See supplementary section 3 for more statistics and discussion.} \textbf{d}. Stray field map of a selected FM area (black dashed rectangle) in (b), \textcolor{black}{after a smooth polynomial background subtraction}, showing {\color{black}correlated} field variation features within FM domains. \textcolor{black}{\textbf{e}. 2D autocorrelation of the field map of (d), revealing hexagonal textures with {\color{black}correlation length} in the long edges around a=340 nm, exceeding the Moiré magnetic periodicity of 36.4 nm at a 1.1° twist angle, giving $a/a_{M}\approx 9.3$. The short edge gives $a/a_{M}\approx 6.2$.}}
  \label{fig:figure2}
\end{figure}

In addition to the "uniform" FM domain shown in Fig. \ref{fig:figure2}a and \ref{fig:figure2}b, we noticed small varying features inside these domains, which are shielded behind a strong stray field gradient generated from the FM-to-AFM domain boundaries. {\color{black} By applying a 3rd order polynomial background subtraction to eliminate the domain boundary field for the magnetic texture in Fig. \ref{fig:figure2}b, we observed weak textures of about 100 $\mu\text{T}$, as shown in Fig. \ref{fig:figure2}d (more details in Supplementary section 5).} To confirm the underlying {\color{black}length scale}, the autocorrelation of the 2D stray field map is computed:

\begin{equation}
	AC(\Delta x, \Delta y)=\sum_{x,y} M(x,y)M(x+\Delta x, y + \Delta y)
    \label{eq:ac}
\end{equation}

where $M(x,y)$ is the value of the 2D field map. This serves to reduce random noise caused by disorder and to enhance the underlying {\color{black}correlation} in 2D maps \cite{song2021direct,reith2017analysing}. The autocorrelation of the field map reveals hexagonal magnetic textures with a {\color{black}length scale} reaching 340 nm, {\color{black}9.3 times greater than the expected Moiré magnetic wavelength of 36.4 nm for 1.1° twisted samples, giving $a/a_{M}\approx 9.3$.} Given the measured field strength, we attributed these observed textures to non-collinear AFM orders within one bilayer, coexisting with a uniform FM response in the other bilayer. Additionally, similar weak magnetic textures were also observed in the FM region of 0.5° twisted samples with a {\color{black}length scale} of 173 nm, giving $a/a_{M}\approx 2.4$ (Supplementary section 5). Interestingly, there is an inverse dependence of the magnetic {\color{black}length scale} $a$ on the underlying Moiré wavelength $a_M$.

\begin{figure}[h!]
  \centering
  \includegraphics[width=1\textwidth]{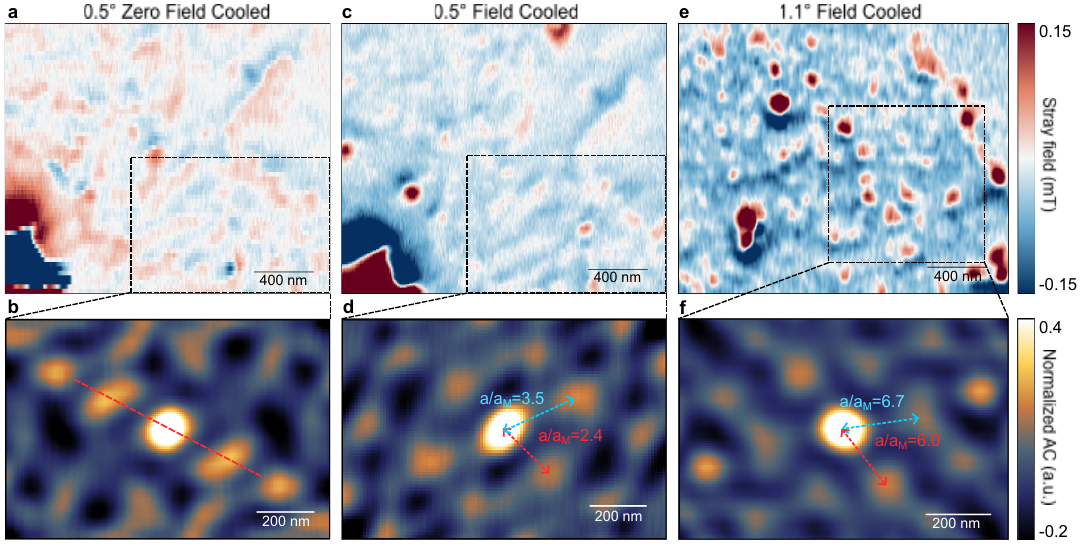}
  \caption{\textbf{a}. A representative stray field map of 0.5° tDB CrI$_3$ after zero field cooldown, showing stripe-like patterns in an AFM region. \textbf{b}. Autocorrelation of a selected area (dashed black rectangle) in (a), showing a 1D {\color{black}correlation} in the diagonal direction, highlighted by the red dashed line. \textbf{c}. Stray field map of the same sample area in (a) after 500 mT-field cooldown, revealing dot-like patterns in the AFM region. \textcolor{black}{\textbf{d}. Autocorrelation of the same area (dashed black rectangle) in (c), showing a hexagonal texture with a spacing of $a/a_M\approx$ 3.5 in the long edge and $a/a_M\approx$ 2.4 in the short edge.} \textbf{e}. Stray field map of 1.1° tDB CrI$_3$ after 500 mT-field cooling,  showing dot-like patterns. \textcolor{black}{\textbf{f}. Autocorrelation of a selected area (dashed black rectangle) in (e), revealing a similar hexagonal pattern with a spacing of $a/a_M\approx$ 6.7 in the long edge and $a/a_M\approx$ 6.0 in the short edge.}}
  \label{fig:figure3}
\end{figure}

A similar analysis can be applied to the AFM domains, as shown in Fig. \ref{fig:figure3}. In this case, no background stray field subtraction is needed. In a 0.5° twisted sample, we observed a weak magnetic response at zero field cooling (ZFC) with stripe-like patterns in the AFM domains (Fig. \ref{fig:figure3}a), suggesting the presence of AFM textures within twisted double bilayer regions. The autocorrelation of a selected area revealed a unilateral {\color{black}correlation} in one direction (Fig. \ref{fig:figure3}b). In contrast, no specific patterns are observed in the natural bilayer and 4-layer regions (Supplementary section 9). To enhance magnetic ordering, we applied field cooling at 0.5 T, a process developed in our previous twisted CrI$_3$ experiments\cite{song2021direct}. The observed AFM stripe-liked pattern transformed into dot-like pattern, as shown in Fig. \ref{fig:figure3}c. Performing autocorrelation on the same selected area revealed a long-range {\color{black}hexagonal correlation} (Fig. \ref{fig:figure3}d), which suggests that the existence of DMI energies in the system contribute to the {\color{black}observed} magnetic textures. {\color{black} We identified a wavelength of about 192 nm, giving $a/a_M \approx$ 2.4.} We refer to this new emergent texture of long-range magnetic order in twisted vdW layers as "{\color{black}super-Moiré magnetic texture}".

We examined AFM patterns at larger twist angles to explore the twist angle dependency. At 1.1°, {\color{black}we measured an even larger super-Moiré {\color{black}textures} of about 218 nm, giving $a/a_M \approx$ 6.0} as shown in Fig. \ref{fig:figure3}e and \ref{fig:figure3}f. To corroborate these observations, more devices at these two angles were fabricated, consistently revealing a larger $a/a_M$ at 1.1° than at 0.5° (Supplementary section 6). A similar trend has also been observed for the super-Moiré patterns in the FM regions of the sample, as discussed in the supplementary material. At angles greater than 1.1°, the expected Moiré periodicity becomes smaller and noncollinear effects diminish, as also shown earlier\cite{xie2023evidence}, leading to no discernible magnetic textures in 2° twisted samples (Supplementary section 10). 

\section*{Super-Moiré Antiferromagnetic Skyrmions}

\begin{figure}[h!]
  \centering
  \includegraphics[width=1\textwidth]{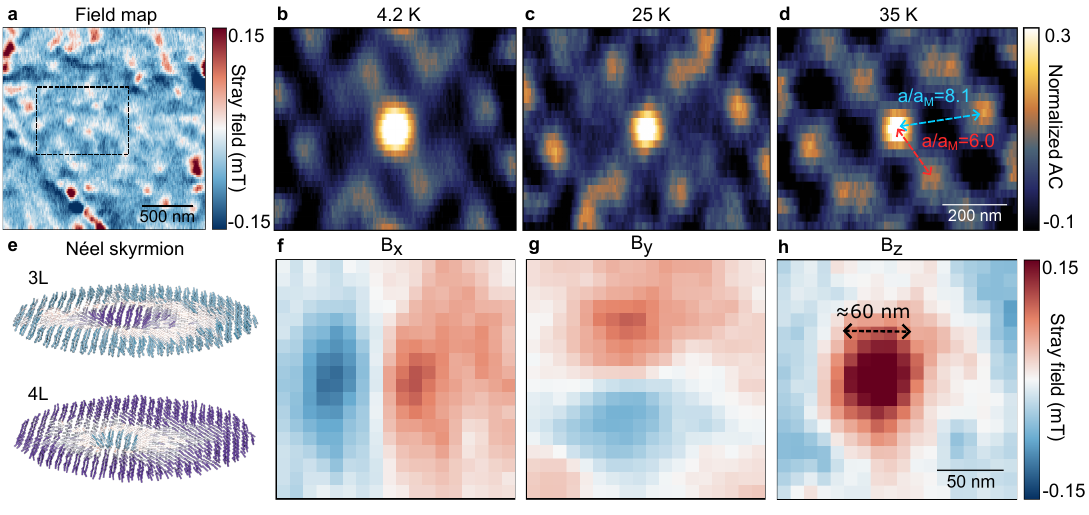}
  \caption{\textbf{a}. A representative stray field map of a 1.1° tDB CrI$_3$ sample after 500 mT field cooldown, taken at 4K, shows dot-like features in the AFM region. An area (dashed black rectangle) was selected to perform autocorrelation at different temperatures. \textbf{b-d}. 2D autocorrelation of the selected area in (a) at 4K, 25K, and 35K, respectively, demonstrating the robustness of the magnetic textures at elevated temperature. The contrast of higher-order peaks becomes more prominent due to the reduction of the critical field as the critical temperature is approached. \textcolor{black}{The spacing of the hexagonal features yields a ratio of $a/a_{M} \approx 8.1$ in the long edges and $a/a_{M} \approx 6.0$ in the short edge.} \textbf{e}. Atomistic simulation of twisted double bilayer CrI$_3$ (tDB CrI$_3$), incorporating interfacial DMI from the hBN/CrI$_3$ interface. Isolated antiferromagnetic (AFM) Néel-type skyrmions emerge in layers 3 and 4. Purple (blue) indicates out-of-plane spins pointing upward (downward), while white denotes in-plane spin orientation. \textbf{f-h}. $B_x$, $B_y$, and $B_z$ stray field reconstruction from fine scan of single dots in (a). The field profile resembles the simulated Néel-type skyrmion shown in (e). From the $B_z$ profile, the feature size is about 60 nm.}
  \label{fig:figure4}
\end{figure}

With temperatures increasing from 4 K to 35 K, these super-Moiré patterns in the twisted sample remain robust, preserving their super-Moiré {\color{black}wavelength} despite changes in temperature, as shown in Fig. \ref{fig:figure4}a-d. The contrast of higher-order peaks became more prominent due to the reduction of the critical field at elevated temperatures. Furthermore, these super-Moiré patterns were robust against magnetic fields, with the features becoming more prominent at higher applied fields (Supplementary section 7). These signatures confirm the emergence of robust noncollinear magnetic textures with AFM orientations. Here, the Moiré lattice and substrate effects can break the intrinsic symmetry of CrI$_3$, introducing Moiré and substrate DMI, and modulate the energy difference between various spin configurations. Under such conditions, the system can favor the formation of hexagonal patterns with skyrmion lattices. A more quantitative analysis of various magnetic energies, including magnetic anisotropy, exchange interaction, dipolar interaction and DMI, is presented in Supplementary Section 19, where isolated skyrmions and other main features such as extended AFM-type magnetic textures can emerge. However, due to the significantly increased complexity for larger simulation domains, the current simulation model\cite{sivadas2018stacking,zheng2023magnetic} cannot incorporate magnetic relaxation and complex lattice relaxation\cite{nam2017lattice,wang2024fractional}, such as magneto-elastic coupling\cite{mcguire2015coupling,guo2021structural,cantos2021layer} and periodic lattice distortions\cite{sung2022torsional}, which are typically observed in actual twisted systems. Such lattice relaxations can also contribute to the formation of super-Moiré skyrmion lattices through the strong modulation of competing energies, requiring further theoretical studies of the lattice relaxation in magnetic systems.

To further confirm the topological nature of the skyrmion, we conducted zoom-in scans on several of these small magnetic features. A representative scan is shown in Fig. \ref{fig:figure4} f-h. The vector magnetic fields ($B_x$, $B_y$, and $B_z$) were derived from the stray field map ($B_{NV}$) detected by the NV probe (See also Supplementary section 2 and 12). Each single dot-like feature exhibits spin configurations consistent with Néel skyrmions\cite{bogdanov2020physical}, which are also observed in the simulation result shown in Fig. \ref{fig:figure4} e. Atomic-scale simulations spanning 450 by 450 nm revealed that isolated Néel skyrmions can form within twisted samples. It is important to note that {\color{black}the skyrmions exhibit variations in size and shape, which we attribute to local strain and Moiré disorder in the twisted samples.} Such skyrmion lattices may contribute to a topological magnetic response, and {\color{black}the same region also exhibits a magnetic signal in RMCD measurements (see Supplementary Fig. S30)}. This, in turn, could account for the observations in earlier RMCD and MOKE measurements, in addition to the net FM domains \cite{kato2023topological,li2024topological}. To the best of our knowledge, our observations represent the first visualization of skyrmion features in twisted 2D systems. 

\section*{Discussion}
In summary, we have utilized state-of-the-art scanning quantum microscopy to explore the nanoscale magnetic responses of tDB CrI$_3$. Moving beyond the conventional framework of Moiré magnetism, we identified super-Moiré antiferromagnetic order at small twist angles (e.g., 0.5°, 1.1°), arising from the interplay between the twisted Moiré potential and complex magnetic competition. More importantly, we uncovered the emergence of super-Moiré {\color{black}magnetic textures} with inverse dependence as the underlying Moiré wavelength, as well as the signatures of Néel-type skyrmions, which have not been reported in any twisted magnetic systems to date. These long-range magnetic textures can also appear in twisted double trilayers (Supplementary section 10) and may even be present in more general twisted devices, reshaping earlier theoretical models \cite{sivadas2018stacking,xiao2021magnetization}. These call for further investigation to unravel the emergence of super-Moiré magnetic textures and other stacking-related behaviors, such as locally enhanced transition temperatures\cite{huang2023revealing,li2024observation} and rotational stacking faults which affect spin ordering\cite{Kwanpyo24}. 

Our observations also enrich the understanding of 2D magnetism, in which strong magnetic anisotropy was believed to be required to stabilize long-range magnetic order, conditions generally considered unfavorable for the formation of magnetic textures such as skyrmion lattices. However, in twisted systems, the Moiré effect introduces intense magnetic competition that not only stabilizes topological spin textures, but also preserves local single-ion anisotropy, thereby reconciling these contradictory requirements. Moreover, this non-dependence of anisotropy for long-range ordering confirms early predictions\cite{Jenkins_2022} where fundamental interactions (e.g., isotropic exchange) are sufficient to create robust magnetism in 2D (Supplementary section 19).    
In addition, our investigations can be generalized to a range of 2D magnetic families  (CrBr$_3$, CrCl$_3$, CrSBr, Fe$_5$GeTe$_2$, etc.) \cite{cai2019atomically,akram2021moire,lee2021magnetic,Balicas23,Koperski24,Yonathan23,Coronado24,Augustin2021,Balicas24} which host very different interaction strengths and symmetries. Moiré engineering provides unconventional tuning knobs to manipulate multiple magnetic interactions with complex electronic competitions to drive the development of skyrmionic devices\cite{fert2017magnetic,chen2024all}. In this context, it may enable the discovery of other exotic magnetic phases such as half-charge merons\cite{gobel2021beyond}, topological magnons\cite{zhang2013topological,mcclarty2022topological}, and quantum spin liquids\cite{takagi2019concept} driven by the Moiré effect.

\clearpage

\bmhead{Acknowledgements}

K.C.W., R.P., S.J., X.Z., M.L., Y.T.K., R.S., and J.W. acknowledge funding support by the EU via the project C-QuEnS 101135359, BMBF via project QCOMP 03ZU1110HA, QSOLID 13N16159 and QMAT 03ZU2110HA, and the Zeiss Foundation through the QPhoton. J.R. acknowledges Ricardo Rama-Eiroa for valuable discussions at the beginning of this project. E.J.G.S. acknowledges computational resources through CIRRUS Tier-2 HPC Service (ec131 Cirrus Project) at EPCC (http://www.cirrus.ac.uk) funded by the University of Edinburgh and EPSRC (EP/P020267/1); and ARCHER2 UK National Supercomputing Service via the UKCP consortium (Project e89) funded by EPSRC grant ref EP/X035891/1. E.J.G.S. acknowledges the EPSRC Open Fellowship (EP/T021578/1) and the Donostia International Physics Center for funding support. \textcolor{black}{A.W.T. acknowledges support from the National Science and Engineering Research Council of Canada (ALLRP/578466-2022) and US Air Force Office of Scientific Research (FA9550-24-1-0360).} E.A. and X.X. at the University of Washington were supported by the Department of Energy, Basic Energy Sciences, Materials Science and Engineering Division (DE-SC0012509). E.A. acknowledges the support by the National Science Foundation Graduate Research Fellowship Program under grant no. DGE-2140004. The device fabrication at U. Washington acknowledges the use of the facilities and instrumentation supported by NSF MRSEC DMR-1719797. Crystal growth at ORNL (M.A.M.) was supported by the US Department of Energy, Office of Science, Basic Energy Sciences, Materials Sciences and Engineering Division.

\bmhead{Author contributions}

J.W., R.P., K.C.W., and X.X. conceived the experiment. K.C.W. conducted the scanning NV measurements, assisted by S.J., M.L., X.Z. and R.P.. E.A. fabricated the tDb CrI$_3$ samples. B.Y., M.C., and A.W.T. provided a comparison tDb sample. M.M., T.T., and K.W. provided 2D crystals. J.R. and E.J.G.S. performed the atomistic simulation and derived the theory with contributions from R.P. and K.C.W.. K.C.W., R.P., J.W., E.A., and X.X. analyzed the experimental results. R.P., K.C.W., E.J.G.S., and J.R. wrote the paper, assisted by Y.T.K.. R.P. and J.W. supervised the research. All authors discussed the results and commented on the paper.





\clearpage
\bibliography{scibib}

\clearpage

\clearpage

\clearpage

\clearpage

\end{document}